\documentclass[12pt]{article}

\begin{document}

\thispagestyle{empty}
\renewcommand{\thefootnote}{\fnsymbol{footnote}}

\begin{flushright}
{\small
FNT/T--95/27\\
November 1995\\}
\end{flushright}

\vspace{1.0cm}

\begin{center}
{\Large\bf   
RANDOM BEAM PROPAGATION IN ACCELERATORS AND STORAGE RINGS}

\vspace{1.0cm}

Stephan I. TZENOV\\

\vspace{1.0cm}

{\it Istituto Nazionale di Fisica Nucleare, Sezione di Pavia,\\
Via Agostino Bassi 6, I-27100 Pavia, ITALY,\\
Electronic mail: TZENOV@AXPPV0.PV.INFN.IT\\

\smallskip

and\\

\smallskip

Institute for Nuclear Research and Nuclear Energy,\\
Bulgarian Academy of Sciences,\\
Blvd. Tzarigradsko Shausse 72, 1784 Sofia, BULGARIA,\\
Electronic mail: TZENOV@BGEARN.ACAD.BG\\}

\end{center}

\vspace{2.5cm}

\begin{center}
{\bf\large   
Abstract }
\end{center}

\begin{quote}
A kinetic equation for the joint probability distribution 
for fixed values of the classical action, momentum and 
density has been derived. Further the hydrodynamic 
equations of continuity and balance of momentum density 
have been transformed into a Schroedinger-like equation,
describing particle motion in an effective electro-magnetic 
field and an expression for the gauge potentials has been 
obtained.
\end{quote}

\newpage


%
\pagestyle{plain}

\section{Introduction}

The dynamics of particles in accelerators and storage rings is 
usually studied on the basis of deterministic causal tools, 
provided by Hamilton's equations of motion. There are, however, 
a number of instances where such a description fails or may not 
be adequate.

The beam circulating in an accelerator may be generally considered 
as an ensemble of nonlinear oscillators. Even in the case when the 
beam is not dominated by space charge these oscillators are weakly 
coupled at least linearly. This coupling is due to direct 
interaction between particles, thus revealing the discrete nature 
of their correlations and/or an interaction between them via the 
surroundings. In an experiment involving macroscopic measuring 
devices the observed quantities are a restricted number of 
variables characterizing the macroscopic state of the beam. Since 
the beam consists of a large number of randomly moving particles 
the macroscopic quantities are subject to deviations around 
certain mean values. These deviations appearing to an observer as 
random events are called fluctuations.

Therefore the particle beam propagates in response to fluctuations, 
automatically implied by the existence of many degrees of freedom. 
Fluctuating contributions remain small in comparison with the 
macroscopic quantities for systems in thermodynamic equilibrium 
\cite{landau}, except at critical points. When a certain dynamic 
(or plasma) instability is encountered, fluctuations are expected 
to grow considerably. As a result the corresponding macroscopic 
evolution exhibits abrupt changes of various thermodynamic 
parameters.

In the light of the above considerations we analyze in the present 
paper the motion of a test particle in the bath provided by the 
rest of the beam. Microscopically, each particle feels the 
fluctuating field (due to electro-magnetic and other possible 
interactions) produced by all the other particles in the beam, and 
therefore it constantly undergoes Brownian motion in phase space. 
The coefficients in the resulting Fokker-Planck equation contain 
the fluctuation spectrum of the interparticle interactions 
\cite{klimontovich, ichimaru}. We are not going to calculate these 
coefficients explicitly here (we hope to do so in a forthcoming 
paper), but rather we study the motion of a test particle suspended 
in a random inhomogeneous medium under the action of external 
forces. The statistical properties of the medium comprising the 
rest of the beam are characterized by a random velocity field 
(which may be regarded also as a fluctuating vector electro-
magnetic potential) and a random potential field. Moreover, we 
consider the beam fluid inviscid. The latter restriction is not 
essential for we presume friction and other sources of dissipation 
(such as synchrotron radiation) could be without effort 
implemented in the development presented here.

Recently a thermal wave model for relativistic charged particle 
beam propagation, building on remarkable analogies between 
particle beam optics and non relativistic quantum mechanics has 
been proposed \cite{fedele}. The conjectured in Reference 4 
Schroedinger-like equation for the transverse motion of 
azimuthally symmetric beams has been derived \cite{tzenov} in 
the framework of Nelson's stochastic mechanics. Further 
development of the model suitable to cover the problem of 
asymmetric beam propagation in accelerators can be found in 
\cite{tzenov1}. In the present paper we recover the Nelson's 
scheme of stochastic mechanics for particle beams from a 
different point of view.

\section{The Model of Random Beam Propagation}

The classical motion of charged particles in an accelerator is 
described usually with respect to a comoving with the beam 
reference frame. Not taking into account chromatic effects the 
dynamics in the longitudinal direction can be decoupled from 
the dynamics in a plane transversal to the orbit. Then the 
evolution of the beam in 6D phase space is governed by the 
Hamiltonian \cite{tzenovd}:

$$
H{\left({\bf x}, {\bf p}; \theta \right)} = 
{\frac {R{\bf p}^2} 2} + {\cal U}
{\left({\bf x}; \theta \right)}, 
\eqno (2.1)
$$

\noindent
where ${\bf x} = {\left( x_1, x_2, x_3 \right)}$, ${\bf p} = 
{\left( p_1, p_2, p_3 \right)}$ and $\theta$ is the azimuthal 
angle, commonly used in accelerator theory as an independent 
variable playing the role of time. The quantity $R$ in equation 
(2.1) denotes the mean radius of the machine. The variables $x_3$, 
$p_3$ constitute a canonical conjugate pair, responsible for the 
longitudinal motion of the beam

$$
x_3 = - {\frac {\sigma} {\sqrt{|{\cal K}|}}} 
sign {\left( {\cal K} \right)} \qquad ; \qquad 
p_3 = - \eta {\sqrt{|{\cal K}|}} 
sign {\left( {\cal K} \right)}, 
\eqno (2.2 )
$$

\noindent
where $\eta$ is the deviation of the actual energy $E$ of the 
particle under consideration with respect to the energy $E_s$ 
of the synchronous particle

$$
\eta = {\frac {1} {\beta_s^2}} {\frac {E-E_s} {E_s}} 
\qquad ; \qquad {\left( \beta_s = {\frac {v_s} {c}}  \right)} 
\eqno (2.3)
$$

\noindent
and $\sigma$ is the displacement of the longitudinal position of 
the particle with respect to the longitudinal position of the 
synchronous particle

$$
\sigma = s - \beta_s ct. 
\eqno (2.4)
$$

\noindent
The quantity ${\cal K}$ is the auto-phasing coefficient (phase slip 
factor), related to the momentum compaction factor $\alpha_M$ 
through the equation

$$
{\cal K} = \alpha_M - {\frac {1} {\gamma_s^2}} \qquad ; \qquad 
{\left( \gamma_s = {\frac {1} {\sqrt{1 - \beta_s^2}}}  \right)}. 
\eqno (2.5)
$$

\noindent
The beam propagation in the plane transversal to the particle orbit 
is described by the canonical conjugate pairs

$$
x_k = {\widetilde x}_k - \eta D_k{\left( \theta \right)} \qquad ; 
\qquad p_k = {\widetilde p}_k - 
{\frac {\eta} {R}} \dot{D}_k{\left( \theta \right)} \qquad 
{\left( k = 1, 2 \right)}. 
\eqno (2.6)
$$

\noindent
In equation (2.6) $\widetilde{x}_k$ is the actual position of our 
particle in the transversal plane and $\widetilde{p}_k$ is the 
canonical conjugate momentum scaled by the total momentum 
$p_s = m_0 \beta_s \gamma_s c$ of the synchronous particle. The 
function $D_k {\left( \theta \right)}$ is the dispersion function 
defined as a solution of the equation

$$
\ddot{D}_k{\left( \theta \right)} + G_k{\left( \theta \right)} 
D_k{\left( \theta \right)} = R^2 K_k{\left( \theta \right)}, 
\eqno (2.7)
$$

\noindent
where $G_k{\left( \theta \right)}$ are the focusing strengths of 
the linear machine in the two transverse directions, 
$K_k{\left( \theta \right)}$ is the local curvature of the orbit 
and the dot [as well as in equation (2.6)] stands for 
differentiation with respect to $\theta$.

The potential function ${\cal U}{\left( {\bf x}; \theta \right)}$ 
in equation (2.1) consists of two parts:

$$
{\cal U}{\left( {\bf x}; \theta \right)} = 
{\cal U}_b{\left( x_1, x_2; \theta \right)} + 
{\cal U}_s{\left( x_3; \theta \right)}, 
\eqno (2.8)
$$

\noindent
where ${\cal U}_b{\left( x_1, x_2; \theta \right)}$ describes the 
transverse motion (betatron motion) and is given by

$$
{\cal U}_b{\left( x_1, x_2; \theta \right)} = 
{\frac {1} {2R}}{\left[G_1{\left( \theta \right)} x_1^2 + 
G_2{\left( \theta \right)} x_2^2 \right]} + 
{\cal V}{\left( x_1, x_2; \theta \right)}, 
\eqno (2.9)
$$

\noindent
while ${\cal U}_s{\left( x_3; \theta \right)}$ is responsible for 
the longitudinal motion (synchrotron motion) and has the form:

$$
{\cal U}_s{\left( x_3; \theta \right)} = 
-sign{\left( {\cal K} \right)} 
{\frac {\Delta E_0} {\beta_s E_s}}
{\frac {c} {2 \pi \omega}} {\cos} 
{\left[ {\frac {\omega {\sqrt{|{\cal K}|}} 
sign{\left( {\cal K} \right)}} {\beta_s c}} x_3 + 
\varphi_0 \right]}. 
\eqno (2.10)
$$

\noindent
In formula (2.10) $\Delta E_0$ is the maximum energy gain per turn, 
$\omega$ and $\varphi_0$ being the angular frequency and phase of 
the accelerating voltage respectively.

From classical mechanics it is well-known that the Hamilton-Jacobi 
equation associated with the Hamiltonian (2.1) is

$$
{\frac {\partial S{\left( {\bf x}; \theta \right)}} 
{\partial \theta}} + {\frac {R} {2}} {\bf p}^2 
{\left( {\bf x}; \theta \right)} + 
{\cal U} {\left( {\bf x}; \theta \right)} = 0, 
\eqno (2.11)
$$

\noindent
where

$$
{\bf p} {\left( {\bf x}; \theta \right)} = 
{\bf \nabla} S{\left( {\bf x}; \theta \right)}. 
\eqno (2.12)
$$

\noindent
For a given arbitrary integral of equation (2.11) a family of 
trajectories ${\bf q} {\left( \theta \right)}$ is generated 
that solve the first order (vector) differential equation

$$
\dot{\bf q} {\left( \theta \right)} = 
R {\bf p} {\left[ {\bf q} {\left( \theta \right)}; 
\theta \right]}. 
\eqno (2.13)
$$

\noindent
Moreover the continuous distribution of trajectories with 
associated density $\varrho {\left( {\bf x}; \theta \right)}$ 
obeys the continuity equation

$$
{\frac {\partial \varrho {\left( {\bf x}; \theta \right)}} 
{\partial \theta}} + R {\bf \nabla} \cdot 
{\left[ \varrho {\left( {\bf x}; \theta \right)} 
{\bf p} {\left( {\bf x}; \theta \right)}
\right]} = 0 
\eqno (2.14)
$$

\noindent
and in addition, taking the gradient of equation (2.11) we 
obtain an equation for ${\bf p} {\left( {\bf x}; \theta \right)}$

$$
{\frac {\partial {\bf p}} {\partial \theta}} + 
R {\left( {\bf p} \cdot {\bf \nabla} \right)} 
{\bf p} + {\bf \nabla} 
{\cal U} {\left( {\bf x}; \theta \right)} = 0. 
\eqno (2.15)
$$

\noindent
Thus the system (2.14), (2. 15) [or equivalently (2.11) and 
(2.14)] represents a closed set of equations, describing the 
Hamilton-Jacobi fluid as a mechanical system living in 
configuration space \cite{guerra}.

Let us consider the motion of a test particle in the fluid 
comprised by the rest of the beam. No dissipative forces of 
Stokes type are present, as soon as we assume the beam fluid 
to be inviscid. However, the discrete nature of collisions 
between particles (intra-beam scattering) give rise to kinetic 
fluctuations in the one particle distribution function. As a 
consequence the gas-dynamic functions 
$\varrho{\left( {\bf x}; \theta \right)}$ and 
${\bf p}{\left( {\bf x}; \theta \right)}$ of the Hamilton-Jacobi 
fluid fluctuate as well. Fluctuating contributions to the 
one particle distribution function are generated also by the 
electro-magnetic interaction between particles in the beam. All 
this means that the beam is a real medium with a finite number 
of particles within a physically infinitesimal volume and 
substituting it by a continuous medium is not justified 
\cite{klimontovich}. External noise could be introduced into 
the beam from the surroundings (RF noise, fluctuations in the 
parameters of magnetic elements, etc.), which complexifies the 
physical picture additionally.

Leaving more speculations aside we consider the motion of our 
test particle in a random inhomogeneous medium and random 
velocity field. The particle dynamics is governed by the 
Hamiltonian

$$
H{\left({\bf x}, {\bf p}; \theta \right)} = 
{\frac {R{\bf p}^2} 2} + {\bf p} \cdot {\bf z}
{\left( \theta \right)} + 
{\cal U}{\left({\bf x}; \theta \right)} + 
{\widetilde {\cal U}}{\left({\bf x}; \theta \right)}, 
\eqno (2.16)
$$

\noindent
where ${\bf z}{\left( \theta \right)}$ is a random velocity 
field with formal correlation properties

$$
{\left \langle {\bf z}{\left( \theta \right)} \right \rangle} 
= 0 \qquad ; \qquad 
{\left \langle z_k{\left( \theta \right)} 
z_m{\left( \theta_1 \right)} \right \rangle} = 
R \epsilon_{km} \delta {\left( \theta - \theta_1 \right)}. 
\eqno (2.17)
$$

\noindent
The quantity 
${\widetilde {\cal U}}{\left({\bf x}; \theta \right)}$ 
is a random potential accounting for the fluctuation of the 
medium. In what follows we shall consider the random potential 
field ${\widetilde {\cal U}}{\left({\bf x}; \theta \right)}$ 
$\delta$-correlated with zero mean value and correlation 
function $A{\left( {\bf x}, {\bf x}_1; \theta \right)}$

$$
{\left \langle
{\widetilde {\cal U}}{\left({\bf x}; \theta \right)}
\right \rangle} = 0 \qquad ; \qquad 
{\left \langle
{\widetilde {\cal U}}{\left({\bf x}; \theta \right)}
{\widetilde {\cal U}}{\left({\bf x}_1; \theta_1 \right)}
\right \rangle} = 
A{\left( {\bf x}, {\bf x}_1; \theta \right)} 
\delta {\left( \theta - \theta_1 \right)}. 
\eqno (2.18)
$$

\noindent
In the Hamiltonian (2.16) we have discarded the ``constant'' 
term ${\bf z}^2{\left( \theta \right)}/2R$ for it does not 
give contribution to the dynamics. Moreover, $\epsilon_{km}$ 
has the dimension of emittance and we call it thermal beam 
emittance tensor.

We would like to note that the equation of random trajectories 
[replacing now equation (2.13)]

$$
\dot{\bf q} {\left( \theta \right)} = 
R {\bf p} {\left[ {\bf q} {\left( \theta \right)}; 
\theta \right]} + {\bf z}{\left( \theta \right)} 
\eqno (2.19)
$$

\noindent
is in fact the equation for the characteristics of

$$
{\frac {dC} {d \theta}} = 
{\frac {\partial C} {\partial \theta}} + 
{\left( R {\bf p} + {\bf z} \right)} \cdot 
{\bf \nabla} C = 0
\eqno (2.20)
$$

\noindent
describing the mixing of concentrations 
$C{\left( {\bf x}; \theta \right)}$ of different species in 
a random velocity field.

Instead of equations (2.11), (2.14) and (2.15) we have now the 
system

$$
{\frac {\partial S} {\partial \theta}} + 
{\frac {R} {2}} {\bf p}^2  + 
{\bf p} \cdot {\bf z} {\left( \theta \right)} + 
{\cal U} {\left( {\bf x}; \theta \right)} + 
\widetilde {\cal U} {\left( {\bf x}; \theta \right)} = 0, 
\eqno (2.21a)
$$

$$
{\frac {\partial \varrho} {\partial \theta}} + 
{\bf \nabla} \cdot 
{\left \{ {\left[ R {\bf p} + {\bf z} {\left( \theta \right)} 
\right]} \varrho \right \}} = 0, 
\eqno (2.21b)
$$

$$
{\frac {\partial {\bf p}} {\partial \theta}} + 
{\left[ {\left( R {\bf p} + {\bf z} \right)} 
\cdot {\bf \nabla} \right]} {\bf p} + 
{\bf \nabla} {\cal U} {\left( {\bf x}; \theta \right)} + 
{\bf \nabla} \widetilde {\cal U} 
{\left( {\bf x}; \theta \right)} = 0, 
\eqno (2.21c)
$$

\noindent
which specifies the evolution law of the Hamilton-Jacobi fluid 
with hydrodynamic Langevin sources.

\section{Kinetic Equation for the One-Point Probability Density}

We define a joint probability density for fixed values of the 
classical action $S{\left( {\bf x}; \theta \right)}$, of the 
momentum ${\bf p}{\left( {\bf x}; \theta \right)}$ and the 
density of random trajectories 
$\varrho {\left( {\bf x}; \theta \right)}$ as

$$
W{\left( S, {\bf p}, \varrho | {\bf x}; \theta \right)} = 
{\left \langle W_r{\left( S, {\bf p}, \varrho | 
{\bf x}; \theta \right)} 
\right \rangle}_{{\bf z}, \widetilde {\cal U}}, 
\eqno (3.1)
$$

\noindent
where ${\left \langle ... \right \rangle}_{{\bf z}, 
\widetilde {\cal U}}$ denotes statistical average over the 
ensemble of realizations of the stochastic processes indicated.

Note that now $S{\left( {\bf x}; \theta \right)}$, 
${\bf p}{\left( {\bf x}; \theta \right)}$ and 
$\varrho {\left( {\bf x}; \theta \right)}$ are random 
functions [more precisely, functionals of the random velocity 
field ${\bf z} {\left( \theta \right)}$ and the random potential 
$\widetilde {\cal U}{\left( {\bf x}; \theta \right)}$] according 
to the system (2.21). A closed kinetic equation for the one-point 
probability density taking into account the gas-dynamic equations 
(2.21) can be derived by particular choice of the dependence of 
$W_r$ on the density $\varrho {\left( {\bf x}; \theta \right)}$ 
\cite{klyatskin}, that is

$$
W_r{\left( S, {\bf p}, \varrho | {\bf x}; \theta \right)} = 
\varrho {\left( {\bf x}; \theta \right)} 
\delta {\left[ S{\left( {\bf x}; \theta \right)} - S \right]} 
\delta {\left[ {\bf p}{\left( {\bf x}; \theta \right)} - 
{\bf p}  \right]}. 
\eqno (3.2)
$$

\noindent
Differentiating equation (3.1) with respect to ``time'' $\theta$ 
and using the gas-dynamic equations (2.21) it is straightforward to 
obtain the following kinetic equation

$$
{\frac {\partial W} {\partial \theta}} + 
R {\bf p} \cdot {\bf \nabla} W + 
{\left( {\frac {R {\bf p}^2} {2}} - {\cal U}  \right)} 
{\frac {\partial W} {\partial S}} - 
{\bf \nabla}{\cal U} \cdot {\bf \nabla}_p W = 
$$

$$
= - {\bf \nabla} \cdot {\left \langle {\bf z} W_r 
\right \rangle} + 
{\frac {\partial} {\partial S}} {\left \langle {\widetilde 
{\cal U}} W_r \right \rangle} + 
{\bf \nabla}_p \cdot {\left \langle W_r {\bf \nabla} 
{\widetilde {\cal U}}  \right \rangle}. 
\eqno (3.3)
$$

\noindent
It is worthwhile to note that if we let $W_r$ depend on 
$\varrho {\left( {\bf x}; \theta \right)}$ through a generic 
function it will turn out that the only possibility to cancel 
terms proportional to ${\bf \nabla} \cdot {\bf p}$ appearing 
in equation (3.3) is to allow $W_r$ be a linear function of 
$\varrho {\left( {\bf x}; \theta \right)}$. However, the 
kinetic equation for the one-point probability density (3.1) 
with an arbitrary dependence on the density of random 
trajectories can be found in a closed form if the Hessian 
matrix

$$
{\cal H}_{mn} = 
{\frac {\partial^2 S{\left( {\bf x}; \theta \right)}} 
{\partial x_m \partial x_n}} 
\eqno (3.4)
$$

\noindent
of the classical action is included in the joint probability 
density \cite{malakhov1}, \cite{malakhov2} and the system 
(2.21) is supplemented with an equation for the quantity 
defined by equation (3.4).

We still have not reached our final goal, since the right hand 
side of equation (3.3) contains the yet unknown correlators of 
the random velocity field, the random potential field and $W_r$. 
In order to split the above mentioned correlations let us 
consider a generic functional ${\cal R} {\left[ F \right]}$ of 
the random Gaussian tensor field $F_{k_1,...,k_n}{\left( 
{\bf r}; \theta \right)}$. Then the following relation holds 
\cite{klyatskin} - \cite{novikov}

$$
{\left \langle F_{k_1,...,k_n} {\left( {\bf r} \right)} 
{\cal R} {\left[ F \right]}  \right \rangle} = 
\int d^n {\bf r}_1  {\left \langle 
F_{k_1,...,k_n} {\left( {\bf r} \right)} 
F_{m_1,...,m_n} {\left( {\bf r}_1 \right)} 
\right \rangle} 
{\left \langle 
{\frac {\delta {\cal R} {\left[ F \right]}} 
{\delta F_{m_1,...,m_n} {\left({\bf r}_1 \right)}}} 
\right \rangle}, 
\eqno (3.5)
$$

\noindent
which is known as the Furutsu-Novikov formula. In (3.5) ${\bf r}$ 
collects all the continuous arguments of the random tensor field, 
$\delta / \delta F_{m_1,...,m_n}{\left( {\bf r} \right)}$ denotes 
the functional derivative with respect to the random field and 
summation over repeated indices is implied. To apply the Furutsu-
Novikov formula (3.5) we need the functional derivatives of 
$S{\left( {\bf x}; \theta \right)}$, 
$\varrho {\left( {\bf x}; \theta \right)}$ and 
${\bf p}{\left( {\bf x}; \theta \right)}$ with respect to the 
random velocity field ${\bf z} {\left( \theta \right)}$ and the 
random potential ${\widetilde {\cal U}} {\left( {\bf x}; 
\theta \right)}$. From equations (2.21a-c) it is easy to find

$$
{\frac {\delta S{\left( {\bf x}; \theta \right)}} 
{\delta z_k {\left( \theta \right)}}} = 
- p_k {\left( {\bf x}; \theta \right)} \quad ; \quad 
{\frac {\delta S{\left( {\bf x}; \theta \right)}} 
{\delta {\widetilde {\cal U}} {\left( {\bf x}_1;\theta \right)}}} = 
- \delta {\left( {\bf x} - {\bf x}_1 \right)} \quad ; \quad 
{\frac {\delta S{\left( {\bf x}; \theta \right)}} 
{\delta {\nabla_{1k}} {\widetilde {\cal U}} 
{\left( {\bf x}_1;\theta \right)}}} = 0, 
\eqno (3.6a)
$$

$$
{\frac {\delta \varrho {\left( {\bf x}; \theta \right)}} 
{\delta z_k {\left( \theta \right)}}} = 
- {\frac {\partial \varrho {\left( {\bf x}; \theta \right)}} 
{\partial x_k}   } \quad ; \quad 
{\frac {\delta \varrho {\left( {\bf x}; \theta \right)}} 
{\delta {\widetilde {\cal U}} {\left( {\bf x}_1; 
\theta \right)}}} = 0 \quad ; \quad 
{\frac {\delta \varrho {\left( {\bf x}; \theta \right)}} 
{\delta \nabla_{1k} {\widetilde {\cal U}} 
{\left( {\bf x}_1; \theta \right)}}} = 0, 
\eqno (3.6b)
$$

$$
{\frac {\delta p_m {\left( {\bf x}; \theta \right)}} 
{\delta z_k {\left( \theta \right)}}} = 
- {\frac {\partial p_m {\left( {\bf x}; \theta \right)}} 
{\partial x_k}} \quad ; \quad 
{\frac {\delta p_m {\left( {\bf x}; \theta \right)}} 
{\delta {\widetilde {\cal U}} {\left( {\bf x}_1; 
\theta \right)}}} = 0 \quad ; \quad 
{\frac {\delta p_m {\left( {\bf x}; \theta \right)}} 
{\delta \nabla_{1k} {\widetilde {\cal U}} 
{\left( {\bf x}_1; \theta \right)}}} = 
- \delta_{km} \delta {\left( {\bf x} - {\bf x}_1 \right)}. 
\eqno (3.6c)
$$

\noindent
By virtue of (3.6), (2.17) and (2.18) we cast equation (3.3) 
into the form:

$$
{\frac {\partial W} {\partial \theta}} + 
R {\bf p} \cdot {\bf \nabla} W + 
{\left( {\frac {R {\bf p}^2} {2}} - {\cal U}  \right)} 
{\frac {\partial W} {\partial S}} - 
{\bf \nabla}{\cal U} \cdot {\bf \nabla}_p W = 
$$

$$
= {\frac {R \epsilon_{km}} {2}} {\nabla_k} {\nabla_m} W - 
{\frac {{\cal A}{\left( \theta \right)}} {2}} 
{\frac {\partial^2 W} {\partial S^2}} + 
{\frac {{\cal C}_{km} {\left( \theta \right)}} {2}} 
{\nabla_{p_k}} {\nabla_{p_m}} W, 
\eqno (3.7a)
$$

\noindent
where we have taken into account the expansion of the correlation 
function (2.18):

$$
A{\left( {\bf x}, {\bf y}; \theta \right)} = 
{\cal A} {\left( \theta \right)} + 
{\cal B}_k {\left( \theta \right)} {\left( x_k - y_k \right)} + 
{\frac {1} {2}} {\cal C}_{km} {\left( \theta \right)} 
{\left( x_k - y_k \right)} {\left( x_m - y_m \right)} + \cdots. 
\eqno (3.8)
$$

\noindent
Without loss of generality the first term in the Taylor expansion 
(3.8) of the correlation function can be taken equal to zero, since 
it does not contribute to the dynamics (it embeds the gauge 
properties of the random potential field and can be scaled to zero). 
Thus we finally arrive at the desired kinetic equation for the 
one- point probability density:

$$
{\frac {\partial W} {\partial \theta}} + 
R {\bf p} \cdot {\bf \nabla} W + 
{\left( {\frac {R {\bf p}^2} {2}} - {\cal U}  \right)} 
{\frac {\partial W} {\partial S}} - 
{\bf \nabla}{\cal U} \cdot {\bf \nabla}_p W = 
$$

$$
= {\frac {R \epsilon_{km}} {2}} {\nabla_k} {\nabla_m} W + 
{\frac {{\cal C}_{km} {\left( \theta \right)}} {2}} 
{\nabla_{p_k}} {\nabla_{p_m}} W, 
\eqno (3.7)
$$

\noindent
The kinetic equation (3.7) is rather complicated to be solved 
directly, so approximate methods to analyze it should be 
involved. For that purpose let us integrate (3.7) over $S$, 
that is exclude the classical action from consideration. We get

$$
{\frac {\partial w} {\partial \theta}} + 
R {\bf p} \cdot {\bf \nabla} w - 
{\bf \nabla}{\cal U} \cdot {\bf \nabla}_p w = 
{\frac {R \epsilon_{km}} {2}} {\nabla_k} {\nabla_m} w + 
{\frac {{\cal C}_{km} {\left( \theta \right)}} {2}} 
{\nabla_{p_k}} {\nabla_{p_m}} w, 
\eqno (3.9)
$$

\noindent
where

$$
w{\left( {\bf p}, \varrho | {\bf x}; \theta \right)} = 
\int dS W{\left( S, {\bf p}, \varrho | {\bf x}; \theta \right)}. 
\eqno (3.10)
$$

\noindent
If we further integrate equation (3.9) over ${\bf p}$ we obtain

$$
{\frac {\partial {\left \langle \varrho \right \rangle}} 
{\partial \theta}} + {\bf \nabla} \cdot {\left[ 
{\left \langle \varrho \right \rangle} 
{\bf v}_{(+)} \right]} - {\frac {R \epsilon_{km}} {2}} 
\nabla_k \nabla_m {\left \langle \varrho \right \rangle} = 0, 
\eqno (3.11)
$$

\noindent
where

$$
{\left \langle \varrho {\left( {\bf x}; \theta \right)} 
\right \rangle} = \int d {\bf p} 
w{\left( {\bf p}, \varrho | {\bf x}; \theta \right)}, 
\eqno (3.12a)
$$

$$
{\left \langle \varrho {\left( {\bf x}; \theta \right)} 
\right \rangle}
{\bf v}_{(+)} {\left( {\bf x}; \theta \right)} = 
R \int d {\bf p} {\bf p} 
w{\left( {\bf p}, \varrho | {\bf x}; \theta \right)}. 
\eqno (3.12b)
$$

\noindent
Defining the osmotic velocity 
${\bf u}{\left( {\bf x}; \theta \right)}$ according to the 
Fick's law

$$
{\left \langle \varrho {\left( {\bf x}; \theta \right)} 
\right \rangle} u_k {\left( {\bf x}; \theta \right)} = 
- {\frac {R \epsilon_{km}} {2}} \nabla_m 
{\left \langle \varrho {\left( {\bf x}; \theta \right)} 
\right \rangle} 
\eqno (3.13)
$$

\noindent
one can write (3.11) in the form of a continuity equation

$$
{\frac {\partial {\left \langle \varrho \right \rangle}} 
{\partial \theta}} + {\bf \nabla} \cdot {\left( 
{\left \langle \varrho \right \rangle} 
{\bf v} \right)} = 0, 
\eqno (3.14)
$$

\noindent
where

$$
{\bf v} {\left( {\bf x}; \theta \right)} = 
{\bf v}_{(+)} {\left( {\bf x}; \theta \right)} + 
{\bf u} {\left( {\bf x}; \theta \right)}
\eqno (3.15)
$$

\noindent
is the current velocity. Next we introduce the stress tensor 
\cite{klimontovich}

$$
\Pi_{mn} {\left( {\bf x}; \theta \right)} = 
R^2 \int d {\bf p} p_m p_n  
w{\left( {\bf p}, \varrho | {\bf x}; \theta \right)}, 
\eqno (3.16)
$$

\noindent
which consists of two parts

$$
\Pi_{mn} = {\left \langle \varrho \right \rangle} 
v_{(+)m} v_{(+)n} + {\cal G}_{mn}. 
\eqno (3.17)
$$

\noindent
The second term in equation (3.17)

$$
{\cal G}_{mn} {\left( {\bf x}; \theta \right)} = 
\int d {\bf p} {\left[ Rp_m - v_{(+)m} \right]} 
{\left[ Rp_n - v_{(+)n} \right]} 
w{\left( {\bf p}, \varrho | {\bf x}; \theta \right)} 
\eqno (3.18)
$$

\noindent
is called the internal stress tensor. Multiplying the kinetic 
equation (3.9) by $R{\bf p}$ and integrating over ${\bf p}$ we 
obtain the transport equation for the momentum density

$$
{\frac {\partial {\left[ {\left \langle \varrho \right \rangle} 
v_{(+)n} \right]}} {\partial \theta}} + 
\nabla_k {\left[ {\left \langle \varrho \right \rangle} 
v_{(+)k} v_{(+)n} \right]} - {\frac {R \epsilon_{km}} {2}} 
\nabla_k \nabla_m {\left[ {\left \langle \varrho \right \rangle} 
v_{(+)n} \right]} = - R {\left \langle \varrho \right \rangle} 
\nabla_n {\cal U} - \nabla_k {\cal G}_{kn}, 
\eqno (3.19)
$$

\noindent
or in alternative form

$$
{\frac {\partial v_{(+)n}} {\partial \theta}} + 
{\left[ {\bf v}_{(-)} \cdot {\bf \nabla} \right]} v_{(+)n} - 
{\frac {R \epsilon_{km}} {2}} \nabla_k \nabla_m 
v_{(+)n} = - R \nabla_n {\cal U} - 
{\frac {1} {\left \langle \varrho \right \rangle}}
\nabla_k {\cal G}_{kn}, 
\eqno (3.20a)
$$

\noindent
where use has been made of equations (3.16-18) and (3.11), and 
the backward velocity field

$$
{\bf v}_{(-)} {\left( {\bf x}; \theta \right)} = 
{\bf v} {\left( {\bf x}; \theta \right)} + 
{\bf u} {\left( {\bf x}; \theta \right)}
\eqno (3.21)
$$

\noindent
has been introduced. One can immediately recognize in the left 
hand side of equation (3.20a) the mean backward derivative 
\cite{guerra, nelson, blanchard} of the forward velocity

$$
{\cal D}_{(-)} v_{(+)n} {\left( {\bf x}; \theta \right)} = 
- R \nabla_n {\cal U} - 
{\frac {1} {\left \langle \varrho \right \rangle}}
\nabla_k {\cal G}_{kn}. 
\eqno (3.20)
$$

\noindent
Perform now ``time'' inversion transformation in equation (3.20) 
according to the relations \cite{guerra}:

$$
\theta \longrightarrow \theta' = - \theta \qquad ; \qquad 
{\bf x}{\left( \theta \right)} \longrightarrow 
{\bf x}'{\left( \theta' \right)} = {\bf x}{\left( \theta \right)} 
\qquad ; \qquad 
{\bf v}{\left( \theta \right)} \longrightarrow 
{\bf v}'{\left( \theta' \right)} = - {\bf v}{\left( \theta \right)}. 
\eqno (3.22)
$$

\noindent
As a consequence of (3.22) one has

$$
{\frac {\partial} {\partial \theta'}} = 
- {\frac {\partial} {\partial \theta}} \qquad ; \qquad 
{\bf \nabla}_{x'} = {\bf \nabla}_x \qquad ; \qquad 
{\bf \nabla}_{v'} = - {\bf \nabla}_v. 
\eqno (3.23a)
$$

\noindent
In addition the forward and backward velocities and mean 
derivatives are transformed as follows \cite{guerra}

$$
{\bf v}_{(\pm)}{\left( {\bf x}; \theta \right)} 
\longrightarrow 
{\bf v}'_{(\pm)}{\left( {\bf x}'; \theta' \right)} = 
- {\bf v}_{(\mp)}{\left( {\bf x}; \theta \right)}, 
\eqno (3.23b)
$$

$$
{\cal D}_{(\pm)}f'{\left( {\bf x}'; \theta' \right)} = 
- {\cal D}_{(\mp)}f{\left( {\bf x}; \theta \right)}, 
\eqno (3.23c)
$$

\noindent
where $f{\left( {\bf x}; \theta \right)}$ is a generic function. 
Since the internal stress tensor ${\cal G}_{kn}$ is a dynamic 
characteristic of motion under time inversion its divergence 
changes sign. This also follows from the particular form of the 
``collision integral'' [the right hand side of the kinetic 
equation (3.7)]. Therefore from (3.20) with (3.22) and (3.23) in 
hand we obtain

$$
{\cal D}_{(+)} v_{(-)n} {\left( {\bf x}; \theta \right)} = 
- R \nabla_n {\cal U} + 
{\frac {1} {\left \langle \varrho \right \rangle}}
\nabla_k {\cal G}_{kn}. 
\eqno (3.24)
$$

Equations (3.20) and (3.24) provide us two opportunities. First, 
summing them up we express the transport equation for the 
momentum density in terms of the current and osmotic velocities 
as

$$
{\frac {\partial {\bf v}} {\partial \theta}} + 
{\left( {\bf v} \cdot {\bf \nabla} \right)}{\bf v} = 
- R {\bf \nabla}{\cal U} + 
{\left( {\bf u} \cdot {\bf \nabla} \right)}{\bf u} - 
{\frac {R \epsilon_{km}} {2}} \nabla_k \nabla_m {\bf u}. 
\eqno (3.25)
$$

\noindent
This is nothing else but the Nelson's stochastic generalization 
of Newton's law \cite{guerra, nelson, blanchard}. Secondly, 
subtracting equations (3.20) and (3.24) we obtain an equation 
for the internal stress tensor to be determined, that is:

$$
{\frac {\partial u_n} {\partial \theta}} + 
{\left( {\bf v} \cdot {\bf \nabla} \right)} u_n - 
{\left( {\bf u} \cdot {\bf \nabla} \right)} v_n + 
{\frac {R \epsilon_{km}} {2}} \nabla_k \nabla_m v_n = 
{\frac {1} {\left \langle \varrho \right \rangle}} 
\nabla_k {\cal G}_{kn}. 
\eqno (3.26)
$$

\noindent
In the isotropic case $\epsilon_{km} = \epsilon \delta_{km}$ 
(see the next section) by virtue of the equation

$$
{\frac {\partial {\bf u}_1} {\partial \theta}} + 
{\bf \nabla}_1 {\left( {\bf u}_1 \cdot {\bf v}_1 \right)} = 
{\frac {R \epsilon} {2}} {\bf \nabla}_1 
{\left( {\bf \nabla}_1 \cdot {\bf v}_1 \right)}
\eqno (3.27)
$$

\noindent
relating the current and osmotic velocity (which is a direct 
consequence of Fick's law and the continuity equation) we 
arrive at the following expression for the internal stress 
tensor:

$$
{\cal G}^{(1)}_{kn} = {\frac {R \epsilon} {2}}
{\left \langle \varrho \right \rangle} 
{\left( \nabla_{1k} v_{1n} + \nabla_{1n} v_{1k} \right)}, 
\eqno (3.28)
$$

\noindent
where [compare with equation (4.2a)]

$$
{\cal G}^{(1)}_{kl} = 
{\left( {\widehat{\cal M}} {\widehat{\cal G}} 
{\widehat{\cal M}}^T \right)}_{kl} = 
{\cal M}_{km} {\cal M}_{ln} {\cal G}_{mn}. 
\eqno (3.29)
$$

\noindent
Transforming back (3.28) to the original coordinates we obtain

$$
{\cal G}_{kn} = {\frac {R {\left \langle \varrho 
\right \rangle}} {2}} 
{\left( \epsilon_{km} \nabla_m v_n + 
\epsilon_{nm} \nabla_k v_m \right)}. 
\eqno (3.30)
$$

Resuming the results of the present section it should be 
mentioned that the continuity equation (3.11) and the transport 
equation for the momentum density (3.19) are equivalent to 
Nelson's scheme of stochastic mechanics. Let us also note 
that the characteristics of the fluctuating beam medium, 
embedded in the random potential (2.18) do not enter the 
simple hydrodynamic approximation procedure adopted here up 
to the second moment. It remains, however to analyze the 
corrections to the evolution law of the Madelung fluid by 
taking into account the balance equation for the kinetic 
energy density. This can be done by employing more complete 
closure techniques to accomplish the transition between 
kinetic and hydrodynamic description.

\section{The Schroedinger-Like Equation}

Our starting point is the system of partial differential 
equations (3.13), (3.14) and (3.25) derived in the preceding 
section, which in fact represents the set of equations 
describing the evolution of the Madelung fluid in stochastic 
mechanics \cite{guerra, blanchard}. Following \cite{tzenov1} 
we perform a coordinate transformation

$$
{\bf x}_1 = {\widehat{\cal M}} {\bf x} \qquad \qquad 
{\left( 
x_{1n} = {\cal M}_{nm} x_m 
\right)}, 
\eqno (4.1)
$$

\noindent
such that the transformed emittance tensor

$$
\epsilon'_{kl} = 
{\left( {\widehat{\cal M}} {\widehat{\epsilon}} 
{\widehat{\cal M}}^T \right)}_{kl} = 
{\cal M}_{km} {\cal M}_{ln} {\epsilon}_{mn} 
\eqno (4.2a)
$$

\noindent
is proportional to the unit tensor $\delta_{kl}$

$$
\epsilon'_{kl} = \epsilon \delta_{kl} 
\eqno (4.2b)
$$

\noindent
by a factor $\epsilon$, where $(\cdots)^T$ denotes matrix 
transposition. The scaling factor can be chosen any of the 
eigenvalues $\epsilon_k$ $(k=1,2,3)$ of the original 
emittance tensor $\epsilon_{kl}$. Provided $\epsilon_{kl}$ 
is symmetric, the matrix ${\widehat{\cal M}}$ has the 
following structure

$$
{\widehat{\cal M}} = {\widehat{\cal A}} 
{\widehat{\cal O}}, 
\eqno (4.3)
$$

\noindent
where ${\widehat{\cal O}}$ is an orthogonal matrix and 
${\widehat{\cal A}}$ is the diagonal matrix

$$
{\cal A}_{kl} = {\sqrt{\frac {\epsilon} {\epsilon_k}}} 
\delta_{kl}. 
\eqno (4.4)
$$

\noindent
Furthermore, the transformed current and osmotic velocities 
are \cite{risken}

$$
{\bf v}_1 = {\widehat{\cal M}}{\bf v} \qquad ; \qquad 
{\bf u}_1 = {\widehat{\cal M}}{\bf u}, 
\eqno (4.5a)
$$

\noindent
while the probability density in the new random coordinates 
is

$$
\varrho_1{\left( {\bf x}_1; \theta \right)} = 
{\left| \det {\widehat{\cal M}} \right|}^{-1} 
{\left \langle 
\varrho{\left( {\bf x}; \theta \right)} 
\right \rangle}. 
\eqno (4.5b)
$$

\noindent
Then the transformed equations of stochastic mechanics read as 

$$
{\frac {\partial \varrho_1} {\partial \theta}} + 
{\bf \nabla}_1 \cdot {\left( 
\varrho_1 {\bf v}_1 \right)} = 0, 
\eqno (4.6a)
$$

$$
\varrho_1 {\bf u}_1 = - {\frac {R \epsilon} {2}} 
{\bf \nabla}_1 \varrho_1, 
\eqno (4.6b)
$$

$$
{\frac {\partial {\bf v}_1} {\partial \theta}} + 
{\left( {\bf v}_1 \cdot {\bf \nabla}_1 \right)}{\bf v}_1 = 
- R \epsilon {\bf \nabla}_{\epsilon} {\cal U} + 
{\left( {\bf u}_1 \cdot {\bf \nabla}_1 \right)}{\bf u}_1 - 
{\frac {R \epsilon} {2}} \nabla^2_1  {\bf u}_1, 
\eqno (4.6c)
$$

\noindent
where

$$
{\left( {\bf \nabla}_{\epsilon} \right)}_n = 
{\frac {1} {\epsilon_n}} \nabla_{1n}. 
\eqno (4.7)
$$

\noindent
We are looking now for a Schroedinger-like equation of the 
type

$$
iR \epsilon {\frac {\partial \psi} {\partial \theta}} = 
{\widehat{\bf H}} \psi 
\eqno (4.8)
$$

\noindent
equivalent to the system (4.6) through the well-known de 
Broglie ansatz

$$
\psi {\left( {\bf x}_1; \theta \right)} = 
{\sqrt{\varrho_1 {\left( {\bf x}_1; \theta \right)}}} 
\exp {\left[ {\frac {i} {R \epsilon}} 
{\cal S} {\left( {\bf x}_1; \theta \right)} \right]}, 
\eqno (4.9)
$$

\noindent
where ${\widehat{\bf H}}$ is a second order differential 
operator with known (constant) coefficients in front of the 
second derivatives. The basic requirement we impose on the 
operator ${\widehat{\bf H}}$ is to be Hermitian

$$
\int d{\bf x}_1 \psi^{\ast}_1 {\widehat{\bf H}} \psi_2 = 
\int d{\bf x}_1 \psi_2 {\widehat{\bf H}}^{\ast} 
\psi^{\ast}_1, 
\eqno (4.10)
$$

\noindent
which defines it (as can be easily shown) up to a generic 
scalar and vector functions. Without loss of generality one 
can write

$$
{\widehat{\bf H}} = {\frac {1} {2}} 
{\left[ iR \epsilon {\bf \nabla}_1 + 
{\bf A}{\left( {\bf x}_1; \theta \right)}
\right]}^2 + 
\Phi{\left( {\bf x}_1; \theta \right)}, 
\eqno (4.11)
$$

\noindent
where the vector function 
${\bf A}{\left( {\bf x}_1; \theta \right)}$ and the scalar 
function $\Phi{\left( {\bf x}_1; \theta \right)}$ define some 
effective electro-magnetic field. Substitution of the ansatz 
(4.9) into equation (4.8) followed by separation of terms by 
real and imaginary part yields:

$$
{\bf v}_1 = {\bf \nabla}_1 {\cal S} - {\bf A}, 
\eqno (4.12a)
$$

$$
{\frac {\partial {\bf v}_1} {\partial \theta}} + 
{\left( {\bf v}_1 \cdot {\bf \nabla}_1 \right)}{\bf v}_1 = 
{\bf E} + {\bf v}_1 \times {\bf B} + 
{\left( {\bf u}_1 \cdot {\bf \nabla}_1 \right)}{\bf u}_1 - 
{\frac {R \epsilon} {2}} \nabla^2_1  {\bf u}_1, 
\eqno (4.12b)
$$

\noindent
where \cite{tonnelat}

$$
{\bf E} = -{\bf \nabla}_1 \Phi - {\frac 
{\partial {\bf A}} {\partial \theta}} \qquad ; \qquad 
{\bf B} = {\bf \nabla}_1 \times {\bf A}. 
\eqno (4.13)
$$

\noindent
Comparing equation (4.12b) with equation (4.6c) we 
conclude that the transformed external force 
$-R \epsilon {\bf \nabla}_{\epsilon}{\cal U}$ equals the 
force produced by the effective electro-magnetic field:

$$
-R \epsilon {\bf \nabla}_{\epsilon}{\cal U} = 
{\bf E} + {\bf v}_1 \times {\bf B}. 
\eqno (4.14)
$$

\noindent
In order to find the electro-magnetic potentials 
${\bf A}{\left( {\bf x}_1; \theta \right)}$ and 
$\Phi{\left( {\bf x}_1; \theta \right)}$ we note that the 
Schroedinger-like equation (4.8) is gauge invariant under 
local phase transformation

$$
\psi_1 {\left( {\bf x}_1; \theta \right)} = 
{\psi {\left( {\bf x}_1; \theta \right)}} 
\exp {\left[ {\frac {i} {R \epsilon}} 
{\cal S}_1 {\left( {\bf x}_1; \theta \right)} \right]}. 
\eqno (4.15)
$$

\noindent
This implies that the equation for the new wave function 
$\psi_1 {\left( {\bf x}_1; \theta \right)}$ has the same 
structure as (4.8) with

$$
{\bf A}{\left( {\bf x}_1; \theta \right)} 
\longrightarrow 
{\bf A}_1{\left( {\bf x}_1; \theta \right)} = 
{\bf A}{\left( {\bf x}_1; \theta \right)} + 
{\bf \nabla}_1{\cal S}_1{\left( {\bf x}_1; \theta \right)}, 
\eqno (4.16a)
$$

$$
\Phi{\left( {\bf x}_1; \theta \right)} 
\longrightarrow 
\Phi_1{\left( {\bf x}_1; \theta \right)} = 
\Phi{\left( {\bf x}_1; \theta \right)} - 
{\frac {\partial {\cal S}_1 
{\left( {\bf x}_1; \theta \right)}} {\partial \theta}}. 
\eqno (4.16b)
$$

\noindent
Moreover, equation (4.14) written in the form

$$
-R \epsilon {\bf \nabla}_{\epsilon}{\cal U} + 
{\bf \nabla}_1 \Phi = - {\frac {\partial {\bf A}} 
{\partial \theta}} + 
{\left( {\bf \nabla}_1 {\cal S} - {\bf A} \right)} 
\times {\left( {\bf \nabla}_1 \times {\bf A} \right)} 
\eqno (4.17)
$$

\noindent
is gauge invariant under (4.15) with

$$
{\cal S}{\left( {\bf x}_1; \theta \right)} 
\longrightarrow 
{\cal S}'{\left( {\bf x}_1; \theta \right)} = 
{\cal S}{\left( {\bf x}_1; \theta \right)} + 
{\cal S}_1{\left( {\bf x}_1; \theta \right)}. 
\eqno (4.18)
$$

\noindent
Choosing ${\cal S}_1{\left( {\bf x}_1; \theta \right)} = 
-{\cal S}{\left( {\bf x}_1; \theta \right)}$ we obtain the 
Euler equation

$$
-R \epsilon {\bf \nabla}_{\epsilon}{\cal U} + 
{\bf \nabla}_1 \Phi_1 = - {\frac {\partial {\bf A}_1} 
{\partial \theta}} - {\bf A}_1 \times 
{\left( {\bf \nabla}_1 \times {\bf A}_1 \right)} 
\eqno (4.19)
$$

\noindent
for the gauge electro-magnetic potentials 
${\bf A}_1{\left( {\bf x}_1; \theta \right)}$ and 
$\Phi_1{\left( {\bf x}_1; \theta \right)}$.

According to (2.8) - (2.10) the external potential 
${\cal U}{\left( {\bf x}; \theta \right)}$ entering the 
Hamiltonian (2.1) can be specified as

$$
{\cal U}{\left( {\bf x}; \theta \right)} = 
{\cal U}_0{\left( {\bf x}; \theta \right)} + 
{\cal V}{\left( {\bf x}; \theta \right)}. 
\eqno (4.20)
$$

\noindent
The term ${\cal U}_0{\left( {\bf x}; \theta \right)}$ 
governs the linear motion and can be written in the form

$$
{\cal U}_0{\left( {\bf x}; \theta \right)} = 
{\frac {1} {2}} {\bf x}^T {\widehat{\bf G}} 
{\left( \theta \right)} {\bf x}, 
\eqno (4.21)
$$

\noindent
where the matrix ${\widehat{\bf G}}{\left( \theta \right)}$ 
is symmetric in general, while 
${\cal V}{\left( {\bf x}; \theta \right)}$ is a sum of all 
nonlinear terms. Further, we split the electric potential 
$\Phi_1{\left( {\bf x}_1; \theta \right)}$ into two parts 
according to the relation

$$
\Phi_1{\left( {\bf x}_1; \theta \right)} = 
\Phi_0{\left( {\bf x}_1; \theta \right)} + 
\Phi'_1{\left( {\bf x}_1; \theta \right)}, 
\eqno (4.22)
$$

\noindent
where

$$
\Phi_0{\left( {\bf x}_1; \theta \right)} = 
{\frac {R} {2}} {\bf x}^T_1 
{\widehat{\bf G}}_1 {\left( \theta \right)} {\bf x}_1 
\qquad ; \qquad 
{\widehat{\bf G}}_1 {\left( \theta \right)} = 
{\widehat{\cal M}} 
{\widehat{\bf G}} {\left( \theta \right)} 
{\widehat{\cal M}}^{-1}, 
\eqno (4.23a)
$$

$$
\Phi'_1{\left( {\bf x}_1; \theta \right)} = 
- {\frac {1} {2}} 
{\bf A}^2_1{\left( {\bf x}_1; \theta \right)}.
\eqno (4.23b)
$$

\noindent
Equation (4.19) takes now the form

$$
-R \epsilon {\bf \nabla}_{\epsilon}{\cal V} = 
- {\frac {\partial {\bf A}_1} {\partial \theta}} + 
{\left( {\bf A}_1 \cdot {\bf \nabla}_1 \right)} {\bf A}_1, 
\eqno (4.24)
$$

\noindent
which transformed back to the original coordinates ${\bf x}$ 
reads as

$$
-R {\bf \nabla}{\cal V} = 
- {\frac {\partial {\bf A}'} {\partial \theta}} + 
{\left( {\bf A}' \cdot {\bf \nabla} \right)} {\bf A}' 
\qquad \qquad 
{\left( {\bf A}' = {\widehat{\cal M}} {\bf A}_1 \right)}. 
\eqno (4.25)
$$

\noindent
Equation (4.12a) suggests an alternative interpretation of the 
vector potential ${\bf A}{\left( {\bf x}_1; \theta \right)}$ as 
the vortex part of the current velocity 
${\bf v}_1{\left( {\bf x}_1; \theta \right)}$. Taking into 
account (4.16a) and the particular choice of the gauge phase 
${\cal S}_1{\left( {\bf x}_1; \theta \right)}$ one can expect 
that ${\bf A}'{\left( {\bf x}; \theta \right)}$ will be 
vortex-free in the original coordinates

$$
{\bf A}'{\left( {\bf x}; \theta \right)} = 
- R {\bf \nabla} \varphi {\left( {\bf x}; \theta \right)}, 
\eqno (4.26)
$$

\noindent
where $\varphi {\left( {\bf x}; \theta \right)}$ is the 
velocity potential \cite{batchelor}. The first integral of the 
equation (4.25) is then

$$
{\frac {\partial \varphi {\left( {\bf x}; \theta \right)}} 
{\partial \theta}} + {\frac {R} {2}} 
{\left[ {\bf \nabla \varphi {\left( {\bf x}; \theta \right)}} 
\right]}^2 + {\cal V}{\left( {\bf x}; \theta \right)} = 
g{\left( \theta \right)}. 
\eqno (4.27)
$$

\noindent
Without loss of generality the generic function 
$g{\left( \theta \right)}$ may be set equal to zero as a result 
of the uncertainty in the definition of the velocity potential 
(4.26). Equation (4.27) is noting else but the Hamilton-Jacobi 
equation (2.11) for the ``classical action'' 
$\varphi {\left( {\bf x}; \theta \right)}$, associated with the 
nonlinear part ${\cal V}{\left( {\bf x}; \theta \right)}$ of 
the external potential ${\cal U}{\left( {\bf x}; \theta \right)}$ 
only.

Performing a second [similar to (4.15)] phase transformation 
according to

$$
\psi_2 {\left( {\bf x}_1; \theta \right)} = 
{\psi_1 {\left( {\bf x}_1; \theta \right)}} 
\exp {\left[ {\frac {i} {R \epsilon}} R 
\varphi {\left( {\bf x}_1; \theta \right)} \right]}. 
\eqno (4.28)
$$

\noindent
we obtain the gauge potentials 
${\bf A}_2 {\left( {\bf x}_1; \theta \right)}$ and 
$\Phi_2 {\left( {\bf x}_1; \theta \right)}$ entering the 
Schroedinger equation for the wave function 
$\psi_2 {\left( {\bf x}_1; \theta \right)}$. They are:

$$
{\bf A}_2 {\left( {\bf x}_1; \theta \right)} = 
R {\left( {\widehat{\bf I}} - {\widehat{\cal A}}^2 \right)} 
{\bf \nabla}_1 
\varphi {\left( {\widehat{\cal M}}^{-1} 
{\bf x}_1; \theta \right)}, 
\eqno (4.29a)
$$

$$
\Phi_2 {\left( {\bf x}_1; \theta \right)} = 
\Phi_0 {\left( {\bf x}_1; \theta \right)} + 
R {\cal V} {\left( {\widehat{\cal M}}^{-1} 
{\bf x}_1; \theta \right)} + 
$$

$$
+ {\frac {R^2} {2}} {\left[ {\bf \nabla}_1 
\varphi {\left( {\widehat{\cal M}}^{-1} 
{\bf x}_1; \theta \right)} \right]}^T 
{\widehat{\cal A}}^2 
{\left( {\widehat{\bf I}} - {\widehat{\cal A}}^2 \right)} 
{\bf \nabla}_1 
\varphi {\left( {\widehat{\cal M}}^{-1} 
{\bf x}_1; \theta \right)}. 
\eqno (4.29b)
$$

\noindent
The Schroedinger-like equation called upon to replace equation 
(4.8) reads as

$$
iR \epsilon {\frac {\partial 
\psi_2 {\left( {\bf x}_1; \theta \right)}} 
{\partial \theta}} = 
{\frac {1} {2}} {\left[ iR \epsilon {\bf \nabla}_1 + 
{\bf A}_2 {\left( {\bf x}_1; \theta \right)} 
\right]}^2 \psi_2 {\left( {\bf x}_1; \theta \right)} + 
\Phi_2 {\left( {\bf x}_1; \theta \right)} 
\psi_2 {\left( {\bf x}_1; \theta \right)}. 
\eqno (4.30)
$$

\noindent
Retracing the sequence of phase transformations (4.15) and 
(4.28) we find

$$
\psi {\left( {\bf x}_1; \theta \right)} = 
\psi_2 {\left( {\bf x}_1; \theta \right)} 
{\exp {\left\{ {\frac {i} {R \epsilon}} 
{\left[ 
{\cal S} {\left( {\bf x}_1; \theta \right)} - 
R \varphi {\left( {\widehat{\cal M}}^{-1} 
{\bf x}_1; \theta \right)} 
\right]}
\right\}}}, 
\eqno (4.31a)
$$

$$
{\bf A}_2 {\left( {\bf x}_1; \theta \right)} = 
{\bf A} {\left( {\bf x}_1; \theta \right)} - 
{\bf \nabla}_1 
{\left[ 
{\cal S} {\left( {\bf x}_1; \theta \right)} - 
R \varphi {\left( {\widehat{\cal M}}^{-1} 
{\bf x}_1; \theta \right)} 
\right]}, 
\eqno (4.31b)
$$

$$
\Phi_2 {\left( {\bf x}_1; \theta \right)} = 
\Phi {\left( {\bf x}_1; \theta \right)} + 
{\frac {\partial} {\partial \theta}} 
{\left[ 
{\cal S} {\left( {\bf x}_1; \theta \right)} - 
R \varphi {\left( {\widehat{\cal M}}^{-1} 
{\bf x}_1; \theta \right)} 
\right]}. 
\eqno (4.31c)
$$

\noindent
The relations (4.31) indicate the equivalence of the 
Schroedinger-like equations (4.8) and (4.30) up to a global 
phase transformation, defined by the constant in the 
coordinates and time phase

$$
{\cal C} = 
{\cal S} {\left( {\bf x}_1; \theta \right)} - 
R \varphi {\left( {\widehat{\cal M}}^{-1} 
{\bf x}_1; \theta \right)} = const. 
\eqno (4.32)
$$

The anisotropy of the random velocity field (2.17) reflects 
on the appearance of the gauge electro-magnetic potentials. 
There are two cases in which the vector potential 
${\bf A}_2 {\left( {\bf x}_1; \theta \right)}$ vanishes 
and the scalar potential 
$\Phi_2 {\left( {\bf x}_1; \theta \right)}$ is equal (up to 
a non essential factor $R$) to the external potential 
${\cal U} {\left( {\bf x}_1; \theta \right)}$. The first 
case is when the external potential is the harmonic 
oscillator potential ${\left( {\cal V} = 0 \right)}$ with 
generally time-dependent frequency, while the second is the 
isotropic case 
${\left( \epsilon_{km} = \epsilon \delta_{km} \right)}$.

\section{Concluding Remarks}

In the present paper we have studied the motion of a test 
particle in a random inhomogeneous medium comprised by the 
rest of the beam. As a result of the investigation performed 
we have shown that Nelson's scheme of stochastic mechanics 
for particle beams in the case of zero friction, is 
equivalent  to hydrodynamic approximation in the kinetic 
equation for the one particle distribution function up to 
the second moment. Further, it has been pointed out that the 
hydrodynamic equations of continuity and momentum density 
can be transformed by a change of coordinates and dependent 
variables into a Schroedinger-like equation. Regardless of 
the type of the external forces one need to introduce a gauge 
electro-magnetic field. If the beam constitutes an isotropic 
medium (holding in the case of symmetric beams) the gauge 
vector potential vanishes and as a consequence the scalar 
potential is equal to the potential that accounts for the 
external force.

The gauge transformation (4.16) is the well-known 
transformation in classical electro-magnetic theory 
\cite{tonnelat} introduced by Weyl, indicating a transition to 
alternative electro-magnetic potentials, which sometimes are 
easier to find compared to the original ones. Besides that, 
the transformed potentials define the same electro-magnetic 
field tensor. Taking into account this fact we have found 
the gauge electro-magnetic potentials explicitly, depending 
on the solution of a Hamilton-Jacobi equation for the 
classical motion of the particle in the anharmonic part of 
the external potential.

The beam circulating in an accelerator consists of a large 
number of particles. Obviously, all of them cannot be in the 
same micro-state. As a result the beam itself generates noise, 
which plays a role similar to the role of perturbation in 
stability theory. The essential difference is that here the 
perturbation is produced by the macroscopic system itself 
(in addition to the noise introduced from the surroundings). 
In the present work we have adopted a phenomenological 
approach to describe beam fluctuations. In this connection 
it remains to compute the statistical properties of the 
beam medium in terms of the fluctuation spectrum, which we 
hope to perform in a forthcoming paper.

\subsection*{Acknowledgements}

It is a pleasure to thank F. Illuminati for careful reading 
of the manuscript and for making useful suggestions. I am 
indebted to M. Roncadelli and A. Defendi as well as to R. 
Fedele and G. Miele for helpful discussions concerning the 
subject of the present paper. Special thanks are due to 
Profs. F. Guerra, S. De Martino and S. De Siena for many 
illuminating discussions on various aspects of stochastic 
mechanics.

 

\end{document}